\documentclass[aps,prl,twocolumn,superscriptaddress,showpacs]{revtex4-1}
\usepackage[dvipdfmx]{graphicx}
\usepackage{latexsym}
\usepackage{amsmath}
\usepackage{txfonts}
\usepackage{helvet}
\usepackage{braket}
\usepackage{amscd}

\usepackage{longtable}

\begin{document}

\title{Site-Selective Mott Transition in a Quasi-One-Dimensional Vanadate V$_6$O$_{13}$} 

\author{Yasuhiro Shimizu}
\author{Satoshi Aoyama}
\author{Takaaki Jinno}
\author{Masayuki Itoh}
\affiliation{Department of Physics, Graduate School of Science, Nagoya University, Furo-cho, Chikusa-ku, Nagoya 464-8602, Japan}
\author{Yutaka Ueda}
\altaffiliation[Present address: ]{Toyota Physical and Chemical Research Institute, Nagakute 480-1192, Japan}
\affiliation{Institute for Solid State Physics, University of Tokyo, Kashiwa, Chiba 277-8581, Japan}

\date{\today}

\begin{abstract} 
	The microscopic mechanism of the metal-insulator transition is studied by orbital-resolved $^{51}$V NMR spectroscopy in a prototype of the quasi-one-dimensional system V$_6$O$_{13}$. We uncover that the transition involves a site-selective $d$ orbital order lifting twofold orbital degeneracy in one of the two VO$_6$ chains. The other chain leaves paramagnetic moments on the singly occupied $d_{xy}$ orbital across the transition. The two chains respectively stabilize an orbital-assisted spin-Peierls state and an antiferromagnetic long-range order in the ground state. The site-selective Mott transition may be a source of the anomalous metal and the Mott-Peierls duality. 
\end{abstract}



\maketitle

	The metal-insulator transition (MIT) involves several instabilities in quasi-one-dimensional (quasi-1D) systems \cite{Mott}. A well-known example is a tetragonal (rutile $R$) phase VO$_2$ in which a competition of Mott and Peierls instabilities has been actively debated for decades \cite{Morin, Eyert, Goodenough, Mott1975, Rice, Biermann, Kotlier, Weber}. A low-temperature nonmagnetic insulating (monoclinic $M_1$) phase has a strong dimerization with $d$ orbital ordering along the chain, supporting the orbital-assisted Peierls mechanism of the MIT \cite{Haverkort}. The complex phase diagram including a partially paramagnetic insulating (monoclinic $M_2$) phase under the slight V oxidation \cite{Pouget} or uniaxial stress \cite{Pouget2} shows degenerated metastable states arising from the competition of electron correlation and electron-lattice coupling. The underlying physics remains a fundamental issue for strongly correlated systems and practical applications to the Mott transistor \cite{Nakano}. 

\begin{figure}
	\includegraphics[width=8cm]{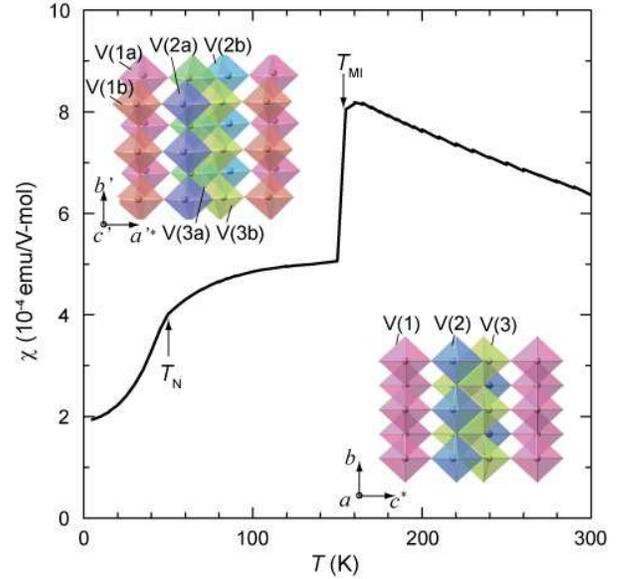}
	\caption{\label{Fig1} 
	Magnetic susceptibility $\chi$ measured with a SQUID magnetometer for a single crystal of V$_6$O$_{13}$ ($H_0 = 1$ T $\parallel b$ and $b^\prime$). 
	The lower and upper insets show crystal structures in the metallic (monoclinic $C2/m$) and insulating (monoclinic $Pc$) phases, respectively \cite{Howing}. 
	}
	\end{figure}

	A Wadsley phase V$_6$O$_{13}$ is another quasi-1D system \cite{Kosuge, Kachi, Kawashima, Dernier, Gossard} in which the microscopic mechanism of the MIT has been debated for half a century. It has been also extensively studied as a potential cathode in a Li-ion battery \cite{West, Saidi, Schmitt2}. There are two frustrated zigzag chains including three vanadium sites, V(1)-V(1) and V(2)-V(3) chains along the $b$ axis, as shown in Fig. \ref{Fig1}. A first-order MIT occurs at $T_{\rm MI}$ = 150 K, accompanied by a drop of the magnetic susceptibility $\chi$ (Fig. \ref{Fig1}) and a weak structural dimerization along the chain \cite{Dernier, Howing}. The lattice symmetry is lowered from a high-temperature monoclinic $C2/m$ phase to a low-temperature monoclinic $Pc$ phase, doubling vanadium sites \cite{Howing}. Whereas the nominal valence is V$^{4.33+} (3d^{0.66})$, the V-O bond valence sum suggests nonuniform valence states V(1)$^{+4.26}$, V(2)$^{+4.92}$, and V(3)$^{+4.61}$ even in the metallic phase \cite{Gossard, Dernier}. Similar to the $M_2$ phase of VO$_2$ \cite{Pouget}, $\chi$ remains paramagnetic in the insulating phase, implying a spin-singlet formation for only a part of the vanadium sites \cite{Itoh, Onoda}. Residual local moments exhibit antiferromagnetic long-range ordering at $T_N$ = 50 K \cite{Ueda, Itoh}.  

	It is noted that the metallic phase of V$_6$O$_{13}$ exhibits several anomalous features. $\chi$ strongly depends on temperature $T$ above $T_{\rm MI}$ (Fig. \ref{Fig1}), implying the presence of local moments. The photoemission spectroscopy shows the depressed density of states at the Fermi level \cite{Shin, Eguchi, Schmitt, Suga}, suggesting that the spectral weights are partially shifted into lower Hubbard bands due to the strongly renormalized electron correlation. These features are in sharp contrast to those of VO$_2$ showing the Pauli paramagnetic susceptibility and the appreciable density of states at the Fermi level \cite{Pouget}, although $T_{\rm MI}$ in V$_6$O$_{13}$ is suppressed to the lower temperature compared to $T_{\rm MI}$ = 342 K in VO$_2$, possibly due to competing ground states. 

	In this Letter, we address the microscopic mechanism of MIT and the origin of the anomalous metal in the quasi-1D vanadate. We focus on the orbital degrees of freedom as a key to understanding the MIT for the $d$ electron system, since the electron correlation can be renormalized into $d$ orbital occupations \cite{Imada}. However, the experimental probe of orbitals has been limited in simple transition-metal compounds \cite{Haverkort}, and the direct observation of the $d$ orbital for several atomic sites remains challenging. Here, we explore NMR spectroscopy as a site-selective orbital probe by utilizing anisotropic hyperfine interactions. We show a dramatic $d$ orbital ordering across $T_{\rm MI}$ for one of the vanadium sites, while remaining as paramagnetic $d_{xy}$ character for the other sites. 

	$^{51}$V NMR measurements were conducted on a single crystal of V$_6$O$_{13}$ made by chemical transport using TeCl$_4$ \cite{Kawashima} and coated with epoxy to prevent crushing at the structural transition. 
	Frequency-swept NMR spectra were obtained from spin-echo signals using a $\pi/2-\tau-\pi/2$ pulse sequence ($\pi/2$ = 1 $\mu$s, $\tau = 5-30$ $\mu$s) with a 0.3 MHz step at a constant magnetic field $H_0 =$ 9.401 T. 
	For a given orientation of the crystal in $H_0$, each vanadium site V($i$) ($i$ = 1, 2, 3 in the metallic phase and $i$ = 1a, 1b, 2a, 2b, 3a, 3b in the insulating phase) gives a set of the $^{51}$V NMR spectra at resonance frequencies $\nu_n^{{\rm V}(i)} = [1 + K^{{\rm V}(i)}]\nu_0 + n\delta \nu^{{\rm V}(i)}$ ($n = -3, -2, ..., 2, 3$), where $K^{{\rm V}(i)}$ are defined as relative shifts from $\nu_0$ = 105.23 MHz and $\delta \nu^{{\rm V}(i)}$ is the nuclear quadrupole splitting frequency for the nuclear spin $I=7/2$. 
	The maximum $|\delta \nu|$ gives the nuclear quadrupole frequency $\nu_Q = 3e^2qQ/h2I(2I - 1)$ with Planck's constant $h$, the nuclear quadrupole moment $eQ$, and the electric field gradient (EFG) $eq = V_{ZZ}$. 
	The nuclear spin-lattice relaxation rate $T_1^{-1}$ was obtained from a single exponential fit to the recovery curve after an inversion $\pi$ pulse at a magic angle ($\delta \nu = 0$) for each V site. 

	\begin{figure}
	\includegraphics[width=8cm]{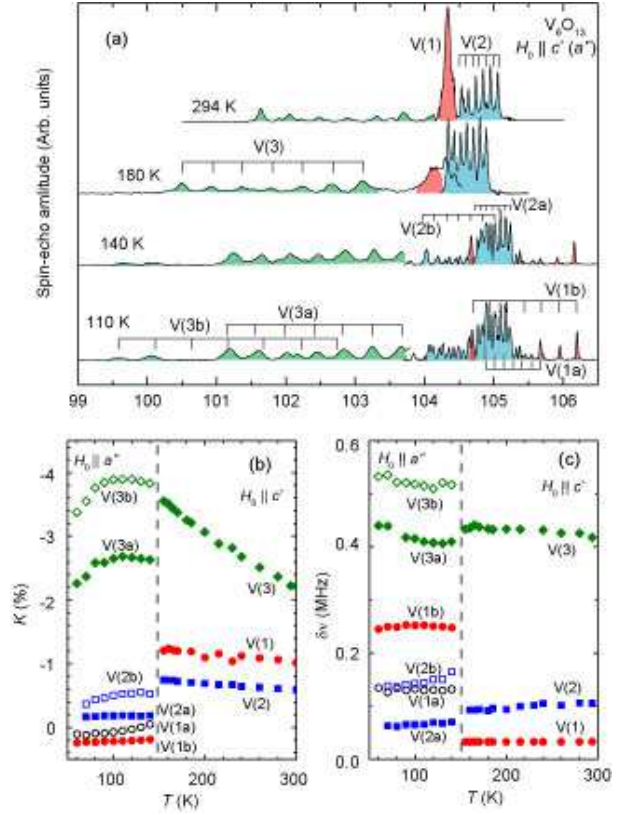}
	\caption{\label{Fig2} 
	(a) $^{51}$V NMR spectra of V$_6$O$_{13}$ [$H_0$ = 9.401 T $\parallel c^* (a^{\prime *})$ for the metallic (insulating) phase], measured with $\tau$ = 6 and $10-30$ $\mu$s for the lower and higher frequency parts, respectively. 
	The red, blue, green, and gray lines are assigned to V(1) [V(1a)], V(2) [V(2a), V(2b)], V(3) [V(3a), V(3b)], and V(1b) based on the angle dependent profiles (Fig. 4). 
	(b) $K$ and (c) $\delta \nu$ obtained from the $^{51}$V NMR spectra. 
	}
	\end{figure}

	Figure \ref{Fig2}(a) shows the $^{51}$V NMR spectra for $H_0 \parallel c^*$ ($a^{\prime *}$). In the metallic phase above 150 K, three sets of the spectrum with the characteristic $K$ and $\delta \nu$ are reasonably assigned to three vanadium sites in reference to the expected valences and the symmetry axes of VO$_6$ distortions. Spike lines located around $\nu_0$ (blue) come from V(2) close to V$^{5+} (3d^0)$ \cite{Itoh}. The others with the negative $K$ and small (red) or large (green) $\delta \nu$ [Figs. \ref{Fig2}(a), (b)] are respectively from V$^{4+}$-like V(1) or V(3) with a weak or strong VO$_6$ distortion. Below $T_{\rm MI}$, the number of spectra increases due to the site doubling. A difference in the V(3a) and V(3b) spectra indicates a slight charge transfer between the vanadium sites across $T_{\rm MI}$. The most noticeable change occurs for the V(1) spectrum: Two sets of the sharp quadrupole split (colored with gray and red) appear with the sizable $\delta \nu$ and vanishing $K$, indicating the $d$ orbital order, as shown below. 
	
 	\begin{figure}
	\includegraphics[width=8cm]{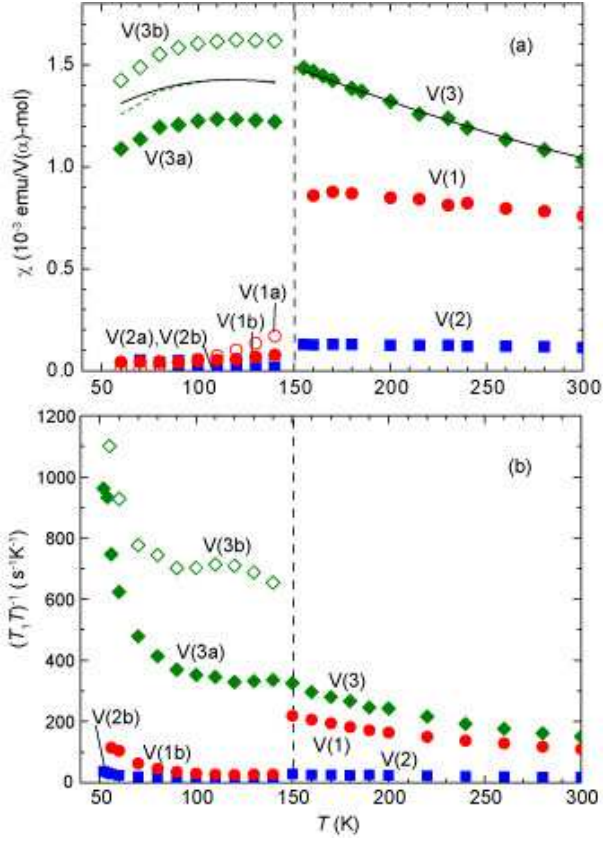}
	\caption{\label{Fig3} 
	(a) Site-dependent local susceptibility obtained from the isotropic $^{51}$V Knight shifts and (b) $1/T_1T$ in V$_6$O$_{13}$. Solid curves are fitting results of the 1D Heisenberg model, whereas the dashed one is an average of $\chi^{\rm V(3a)}$ and $\chi^{\rm V(3b)}$. 
	}
	\end{figure}

	First we focus on the local susceptibility $\chi^{{\rm V}(i)}$ for each vanadium site based on $K$ measurements (Fig. S1, Supplemental Material \cite{SM}). $\chi^{{\rm V}(i)}$ obtained from $K$ is shown in Fig. \ref{Fig3}. $\chi^{\rm V(3)}$ exhibits a strong thermal variation in the metallic phase. It shows a discontinuous increase (decrease) for V(3a) [V(3b)] at $T_{\rm MI}$. The $T$ dependence behaves similar to that of the spin $S=1/2$ 1D Heisenberg antiferromagnet ($\mathcal {H} = \sum J{\bf S}_i\cdot {\bf S}_j$) with an exchange coupling $J = 77$ K for $\chi^{\rm V(3a)}$ and $\chi^{\rm V(3b)}$ below 140 K and with $J = 69$ K for $\chi^{\rm V(3)}$ above 150 K. In contrast, $\chi^{\rm V(1)}$ exhibits rather weak $T$ dependence in the metallic phase and becomes suppressed in the insulating phase, leaving the small Van Vleck susceptibility [$< 5 \times 10^{-4}$ emu/V(1) mol]. The result gives clear evidence for site-selective spin-singlet formations. Namely, the V(1) chain exhibits the MIT from a (Pauli) paramagnetic metal to a nonmagnetic insulator, as seen in VO$_2$ \cite{Pouget}. $\chi^{\rm V(2a/2b)}$ becomes negligible below $T_{\rm MI}$: The on-site $d$ spin is absent on V(2a/2b), rendering V(1a/1b) and V(3a/3b) effectively half-filling. 

	The dynamical property measured with $T_1^{-1}$ is another manifestation of the site-dependent correlation [Fig. \ref{Fig3}(b)]. In contrast to weakly correlated metals with the $T$-linear $1/T_1$, $1/T_1T$ for V(3) increases upon cooling and then becomes further enhanced below $T_{\rm MI}$ for V(3a) and V(3b) due to the evolution of antiferromagnetic fluctuations toward $T_{\rm N}$. For V(1), $1/T_1T$ exhibits a large drop at $T_{\rm MI}$, consistent with the behavior of $\chi^{\rm V(1)}$. Reflecting the dilute spin density, $1/T_1T$ for V(2) is an order of magnitude smaller than those of the other sites. Thus, the site dependence in the spin correlation becomes highlighted below $T_{\rm MI}$. 

	\begin{figure*}
	\includegraphics[width=17cm]{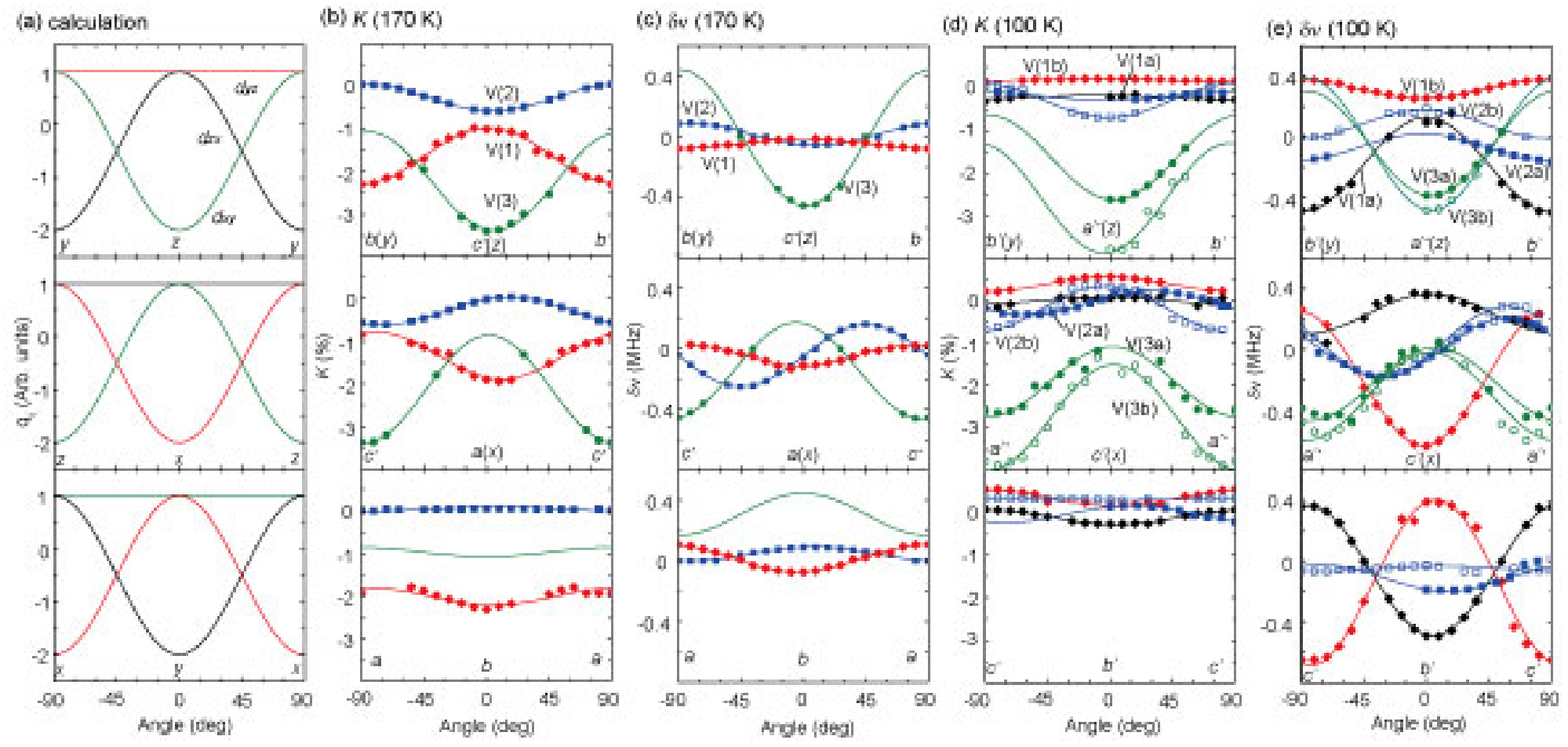}
	\caption{\label{Fig4} 
	(a) Calculated ${\sf q}_{ij}$ proportional to the on-site dipole hyperfine coupling and EFG for $d_{xy}$, $d_{yz}$, and $d_{zx}$ orbitals. 
	The orthogonal coordinate axes $(x, y, z)$ are taken on the VO$_6$ octahedron and parallel to $(a, b, c^*)$ or $(c^\prime, b^\prime, a^{\prime *})$ in Fig. \ref{Fig1}. 
	$(b)-(e)$ Angular dependence of the $^{51}$V Knight shift $K$ and the nuclear quadrupole splitting frequency $\delta \nu$ at 170 and 100 K in V$_6$O$_{13}$. 
	The curves are fitted results with the sinusoidal functions of Volkov's formula \cite{Volkov}. 
	}
	\end{figure*}

	To give deeper insights into the microscopic origin of the MIT, the predominant $d$ orbital occupation is investigated by measuring the angular dependence of $K$ and $\delta \nu $ (Fig. \ref{Fig4}). Since the magnetic susceptibility is isotropic in vanadates, the $K$ and $\delta \nu $ anisotropies are governed by the anisotropic hyperfine couplings expressed with the equivalent operators ${\sf q}_{ij}\equiv \frac{3}{2}(L_i L_j + L_j L_i ) - \delta_{ij}{\bf L}^2$ ($i, j$ = $x, y, z$) \cite{Abragam, Kiyama}, where {\bf L} represents the total orbital angular momentum in the local coordinates $(x, y, z)$ corresponding to $(a, b, c^*)$ or $(c^\prime, b^\prime, a^{\prime *})$ (Supplemental Material \cite{SM}). One can elucidate orbital occupations by extracting ${\sf q}_{ij}$. For a singly occupied $d_{xy}$, $d_{yz}$, or $d_{zx}$ orbital, the components of ${\sf q}_{ij}$ exhibit the cyclic angular dependence [Fig. \ref{Fig4}(a)] with a minimum along the symmetry axis of the orbital. For metals, ${\sf q}_{ij}$ represents the electron cloud distribution for the Hund-coupled bands crossing the Fermi level. With increasing electron correlation, the spin-polarized lower Hubbard band may significantly contribute to the spin susceptibility \cite{Imada}. On the other hand, ${\sf q}_{ij}$ obtained from $\delta \nu $ measures the net $d$ occupation below the Fermi level. 

	As for the metallic phase, $K^{\rm V(3)}$ has a minimum for the $c^*$ axis [Fig. \ref{Fig4}(b)], in good agreement with the $d_{xy}$ anisotropy, whereas $K^{\rm V(1)}$ exhibits out-of-phase angular dependence with a maximum for the $c^*$ axis and the depressed anisotropy. ${\sf q}_{ij}$ is extracted from the spin parts of the axial anisotropy $K_{\rm ax}^s/K_{\rm iso}^s$ and the in-plane asymmetry $K_{\rm an}^s/K_{\rm iso}^s$ expressed with a linear combination of three $t_{2g}$ orbital components $d_{xy}:d_{yz}:d_{zx} = \alpha : \beta : \gamma $ ($\alpha + \beta + \gamma = 1$), where $K_{\rm ax}^{\rm s}= (2K^s_Z - K^s_X - K^s_Y)/3$, $K_{\rm an}^s= (K^s_X - K^s_Y)/2$, and $K_{\rm iso}^s = (K_X^s + K_Y^s + K_Z^s)/3$ (Supplemental Material \cite{SM}). The obtained occupation ratio is listed in Table \ref{table1}. It turns out that V(1) has an admixture of the $d_{yz}$ and $d_{zx}$ components. It is consistent with the depressed value of $\nu_Q$ = 0.1 MHz due to the degenerated orbital occupation (Table S3, Supplemental Material \cite{SM}). The results are also in qualitative agreement with a density functional theory (DFT) calculation ($\alpha : \beta : \gamma= 0.0/0.23/0.38$) \cite{Ohta}. In clear contrast, the V(3) site is mostly polarized to $d_{xy}$, suggesting strong electron correlations under the crystal field. 

\begin{table}[h]
\caption{\label{table1}Axial anisotropy $K_{\rm ax}^{\rm s}/K_{\rm iso}^{\rm s}$ and in-plane asymmetry $K_{\rm an}^{\rm s}/K_{\rm iso}^{\rm s}$, the nuclear quadrupole frequency $\nu_{\rm Q}$, the EFG asymmetry $\eta$, and the orbital occupation in V$_6$O$_{13}$, where the occupations are obtained from the $K$ anisotropy for V($i$) ($i$ = 1, 3, 1a, 1b, 3a, and 3b) and from the nuclear quadrupole frequency for V(1a) and V(1b) (Supplemental Material \cite{SM}).} 
\begin{tabular}{cccccc}
\hline 
site&$K_{\rm ax}^{\rm s}/K_{\rm iso}^{\rm s}$ &$K_{\rm an}^{\rm s}/K_{\rm iso}^{\rm s}$ &$\nu_{\rm Q}$(MHz)&$\eta $&$d_{xy}:d_{yz}:d_{zx}$\\ \hline
V(1) & $-0.38$ &$0.21$&$0.11$& 0.40&0.10:0.56:0.28\\ 
V(3)&$0.77$&$0.03$&$0.46$& 0.26&0.77:0.13:0.10\\ 
\hline
V(1a) & - & - & $0.49$& 0.47&0.11:0.18:0.71\\
V(1b) & - & - & $0.64$& 0.20&0.13:0.78:0.09\\
V(3a)&$ 0.70$&$-0.05$&$0.44$& 0.69&0.74:0.16:0.10\\
V(3b)&$ 0.74 $&$-0.04$&$0.54$& 0.76&0.76:0.14:0.10\\
\hline
\end{tabular}
\end{table}

	On the insulating phase, the angular profiles of $K$ and $\delta \nu$ for V(2a/2b) and V(3a/3b) are similar to those of the metallic phase. It confirms that the orbital states of these sites have less impact on the MIT. The most remarkable change across $T_{\rm MI}$ is the emergence of the huge $\nu_Q$ sites, 0.49 MHz for V(1a) and 0.68 MHz for V(1b) \cite{note}, pointing to $d$ orbital ordering on the V(1a)-V(1b) chain. The minimum of $\delta \nu$ appears along the $b^\prime$ ($c^\prime$) axis for V(1a) [V(1b)], in agreement with the predominant $d_{zx}$ ($d_{yz}$) occupation [Figs. 4 (a), 4(e)]. Namely, the $d_{yz}/d_{zx}$ degeneracy in the metallic phase is completely lifted by the orbital order. The $\delta \nu$ anisotropy gives the largely polarized occupation for V(1a) and V(1b) (Tables \ref{table1} and and Supplemental Material Table S3 \cite{SM}) after subtracting the lattice contribution (Table S2, Supplementa Material \cite{SM}). We thus conclude that the MIT in the present system is attributed to site-selective orbital ordering accompanied by the spin-singlet formation. 

 	\begin{figure}
	\includegraphics[width=8cm]{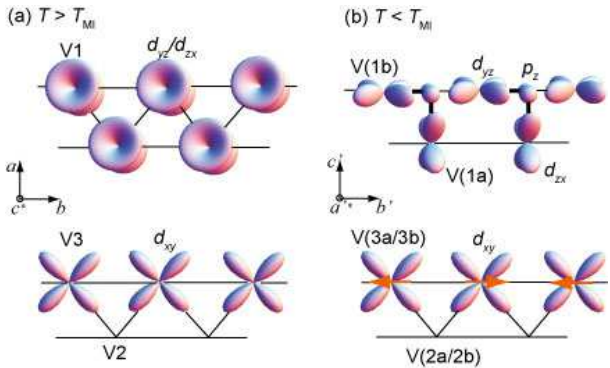}
	\caption{\label{Fig5}
	Configurations of the predominant orbital occupations in the (a) metallic and (b) insulating phases of V$_6$O$_{13}$. The bold lines indicate spin-singlet pairs. The arrows on V(3a)/V(3b) denote anitiferromagnetically ordered spins below 50 K \cite{Itoh}. 
	}
	\end{figure}

	The obtained predominant orbitals are depicted in Fig. \ref{Fig5}. Owing to the dilute spin density on V(2), the $d_{xy}$ orbitals on the V(3) sites are weakly coupled each other on the V(2)-V(3) zigzag chain and carry the local moments even in the metallic phase. The degenerated $d_{yz}/d_{zx}$ orbitals on V(1) form the conducting chain, which are lifted into the singly occupied $d_{yz}$ or $d_{zx}$ state below $T_{\rm MI}$. A weak dimerization along the V(1a)-V(1b) chain produces the spin-singlet pairs. The $\pi$ bonding between orthogonal $d$ orbitals should be rather weak. This feature is in clear contrast to the $d$ orbital order in VO$_2$ where the direct $d$-$d$ $\sigma$-bond stabilizes strong spin-singlet pairing \cite{Haverkort}. To explain the vanishing spin susceptibility with a sizable spin gap ($\Delta > 150$ K) in V$_6$O$_{13}$, one may have to consider significant superexchange interactions through the O $p_z$ orbital. Such interactions help to lift the $d_{yz}/d_{zx}$ degeneracy, analogous to the Jahn-Teller mechanism \cite{Ohta}. In context, the MIT involves the spin-Peierls transition assisted by orbital ordering and electron-phonon coupling. The above observation of the paramagnetic local moment on V(3a/3b) provides evidence for the Mott insulating ground state of V$_6$O$_{13}$ with antiferromagnetic ordering \cite{Itoh}, whereas a theory beyond the DFT $+ U$ band calculation \cite{Ohta} is required to describe paramagnetic Mott insulators. It is remarkable that the orbital occupation and the exchange interaction do not show significant changes for V(3) across $T_{\rm MI}$, keeping the Mott insulating features. The results showing the partially localized vanadium sites are consistent with the depressed spectral weight at the Fermi level and the significant weights well below the Fermi level as observed in the XPS measurements \cite{Eguchi}.

	The site-selective Mott transition has been proposed in rare-earth nickelates based on a DFT + dynamical mean field theory calculation \cite{Park}, while it has not been confirmed experimentally. Our observations of the site-dependent $d$ occupations and spin susceptibilities give microscopic evidence for the site-selective Mott localization. Many transition-metal compounds with inequivalent atomic sites have site-dependent local environments and a potential to the site-selective localization, giving anomalous bad metals with the enhanced resistivity and magnetic susceptibility. Even under the weak crystal field, the emergence of the large orbital polarization in the metallic phase may feature the evolution of the lower Hubbard bands. The orbital-resolved NMR provides the site-selective local probe of the orbital order due to strong mass renormalization and can be applied to extensive materials \cite{Shimizu2}. 
	
	To conclude, we have presented the observation of the site-selective metal-insulator transition, which accounts for the anomalous features in the historical quasi-one-dimensional correlated system V$_6$O$_{13}$. The orbital polarization and local spin susceptibility are elucidated for three vanadium sites with different electron correlations. We uncover that the metal-insulator transition is dominated by orbital ordering and spin-singlet pairing for the itinerant vanadium site with the orbital degeneracy, whereas the magnetic site with the singly occupied orbital remains with paramagnetic moments. Thus the spin-Peierls state and antiferromagnetic order coexist in the insulating ground state. The results demonstrate the competing instabilities and the reasonable ways of symmetry breaking in the frustrated one-dimensional system. 
	
	We thank Y. Ohta for fruitful discussion and S. Inoue for technical assistance. This work was financially supported by KAKENHI (No. 25610093, No. 23225005, No. 24340080, and No. 25610092) from the Japan Society for the Promotion of Science.

\pagebreak

\setcounter{equation}{0}
\setcounter{figure}{0}
\setcounter{table}{0}
\setcounter{page}{1}
\makeatletter
\renewcommand{\theequation}{S\arabic{equation}}
\renewcommand{\thefigure}{S\arabic{figure}}
\renewcommand{\bibnumfmt}[1]{[S#1]}

\begin{center}
\textbf{\large Supplemental Material for Site-Selective Mott Transition in a Quasi-One-Dimensional Vanadate V$_6$O$_{13}$}
\end{center}
	In Supplemental Material, we describe the procedure to obtain the site-dependent local susceptibility and the $d$ orbital occupation from the $^{51}$V NMR measurements in V$_6$O$_{13}$. 

\section{A. Local susceptibility}
	\begin{figure}
	\includegraphics[width=8cm]{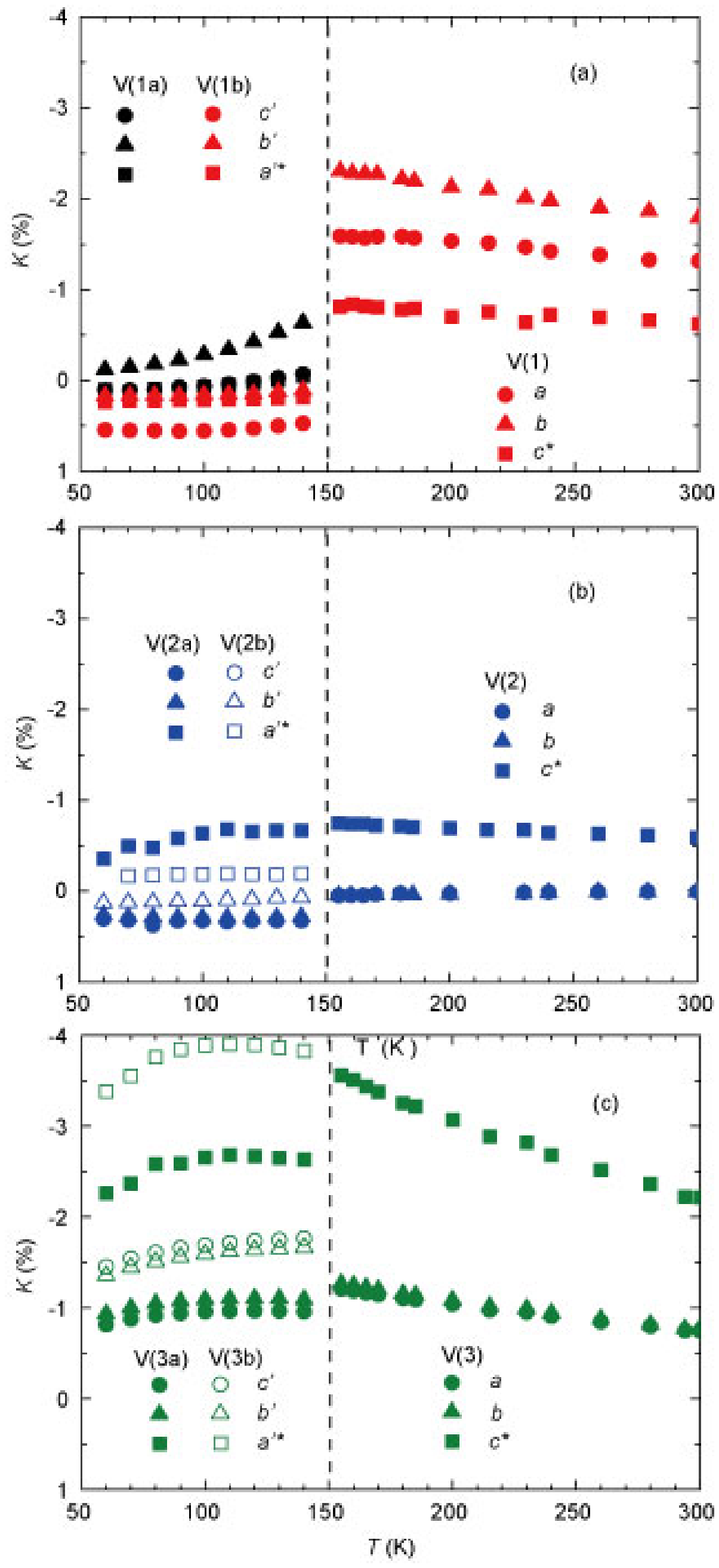}
	\caption{\label{FigS1} 
	Temperature dependence of $K$ obtained from the $^{51}$V NMR spectra for $H_0$ parallel to the three crystal axes in V$_6$O$_{13}$. 
	}
	\end{figure}

	The net magnetic susceptibility $\chi$ (emu/V-mol) observed by the bulk measurement is expressed as an average of the local susceptibilities $\chi^{{\rm V}(i)}$ [emu/V($i$)-mol], namely, 
	\begin{eqnarray}
	\chi = \frac{1}{3}\sum_i \chi^{{\rm V}(i)} 
	\end{eqnarray} 
for the metallic phase ($i$ = 1, 2, 3), and
	\begin{eqnarray}
	\chi = \frac{1}{6}\sum_i \chi^{{\rm V}(i)} 
	\end{eqnarray} 
for the insulating phase ($i$ = 1a, 1b, 2a, 2b, 3a, 3b) of V$_6$O$_{13}$. 
	$\chi^{{\rm V}(i)}$ consists of the spin and orbital components, $\chi_{\rm s}^{{\rm V}(i)}$ and $\chi_{\rm orb}^{{\rm V}(i)}$, for $d$ electrons and the diamagnetic one for core electrons $\chi_{\rm dia}$, 
	\begin{eqnarray}	
	\chi^{\rm V(i)} = \chi_{\rm s}^{{\rm V}(i)} + \chi_{\rm orb}^{{\rm V}(i)} + \chi_{\rm dia}.
	\end{eqnarray} 
	For $3d$ compounds, $\chi_{\rm orb}$ is governed by the temperature-independent Van-Vleck paramagnetic susceptibility $\chi_{\rm VV}$. 
	We redefine $\chi$ after subtracting $\chi_{\rm dia} = -6 \times 10^{-6}$ emu/V-mol. 
	
	NMR probes $\chi^{{\rm V}(i)}$ as the isotropic Knight shift $K_{\rm iso}$ consisting of the spin part 
\begin{eqnarray}
K_{\rm iso}^{\rm s} = A^{\rm s}\chi_{\rm s}/N_{\rm A}\mu_{\rm B} 	\end{eqnarray} 
and the orbital part 
\begin{eqnarray}
K^{\rm orb} = A^{\rm orb}\chi_{\rm orb}/N_{\rm A}\mu_{\rm B}
\end{eqnarray} 
with the spin and orbital hyperfine coupling constants, $A^{\rm s}$ and $A^{\rm orb}$, the Avogadro number $N_{\rm A}$, and the Bohr magneton $\mu_{\rm B}$. 
	For vanadates, $A^{\rm s}$ is usually negative due to the predominant core polarization interaction. 
	
	We measured the Knight shift for the three orthogonal directions, ($a$, $b$, $c^*$) in the metallic phase, corresponding to ($c^\prime$, $b^\prime$, $a^{\prime *}$) in the insulating phase, as displayed in Fig. S1. 
	Since $\chi$ would be isotropic for vanadates due to weak spin-orbit coupling, the anisotropy of $K$ arises from the anisotropic dipole hyperfine coupling. 
	The isotropic part, the average of the three orthogonal components, $K_{\rm iso} = (K_a + K_b + K_{c^*})/3$, gives $\chi^{{\rm V}(i)}$ by determining $A^{\rm s}$, as mentioned below. 

\begin{figure}
	\includegraphics[width=8cm]{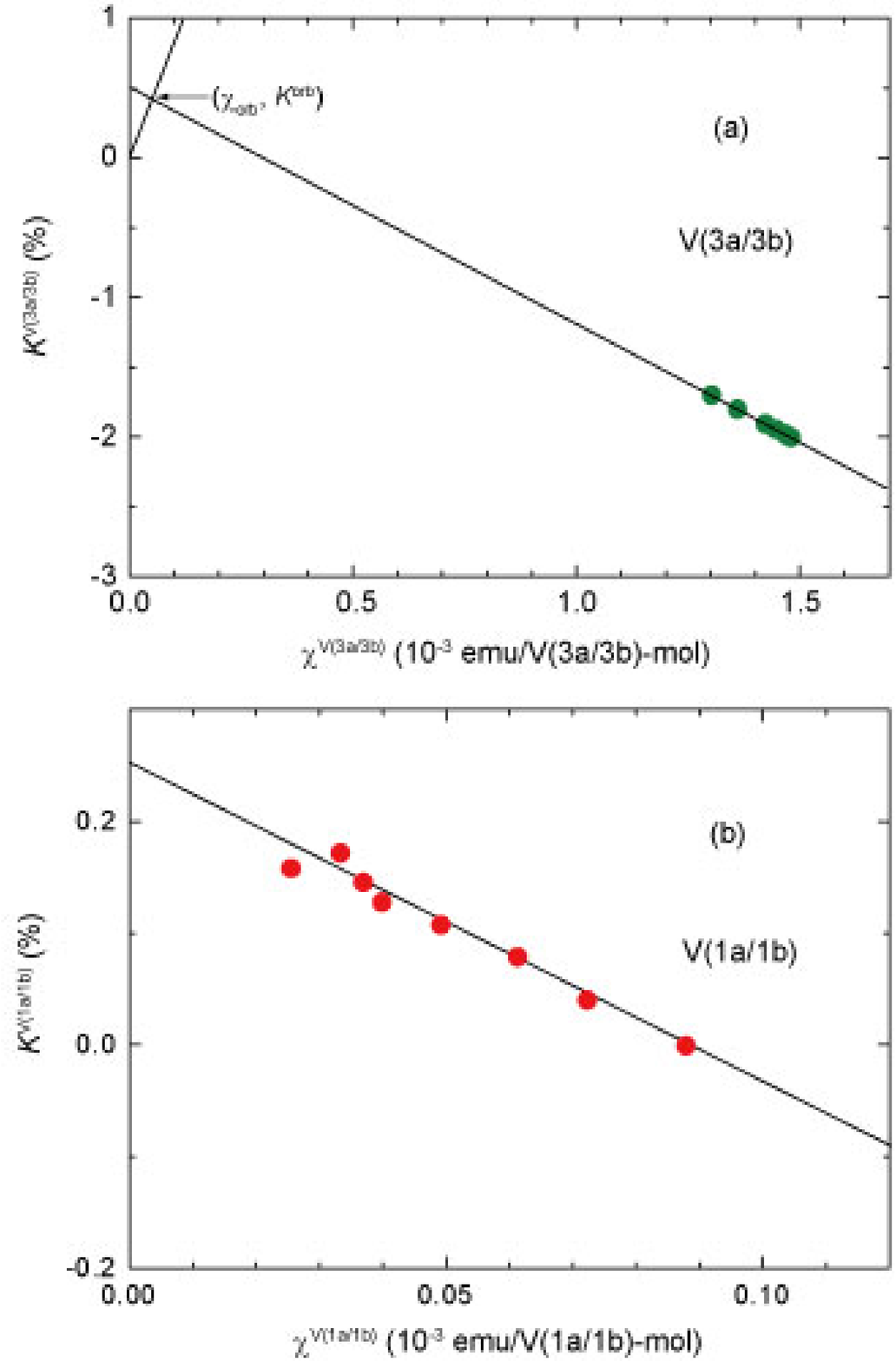}
	\caption{\label{FigS2} 
	$K$ vs $\chi^{{\rm V}(i)}$ plots for (a) V(3a/3b) and (b) V(1a/1b) in V$_6$O$_{13}$.  
	}
	\end{figure}

	In the insulating phase of V$_6$O$_{13}$, $\chi_{\rm s}^{\rm V(1a/1b)}$ and $\chi^{\rm V(2a/2b)}$ go to zero at low temperatures. 
	Then $\chi$ consists of $\chi_{\rm orb}^{\rm V(1a)}$, $\chi_{\rm orb}^{\rm V(1b)}$, and $\chi^{\rm V(3a/3b)}$, where $\chi^{\rm V(3a/3b)} = [\chi_{\rm orb}^{\rm V(3a)} + \chi_{\rm orb}^{\rm V(3b)} + \chi_{\rm s}^{\rm V(3a)} + \chi_{\rm s}^{\rm V(3b)}]/2$. 
	From the Knight shifts at 60 K, $K_{\rm iso}^{\rm V(1a)} = 0.03\%$ and $K_{\rm iso}^{\rm V(1b)} = 0.31\%$, we obtained $\chi_{\rm orb}^{\rm V(1a)} = 4 \times 10^{-6}$ emu/V(1a)-mol and $\chi_{\rm orb}^{\rm V(1b)} = 4 \times 10^{-5}$ emu/V(1b)-mol by using $A^{\rm orb} =2\braket{r^{-3}}$ = 461 kOe/$\mu_{\rm B}$ for V$^{4+}$ (Ref. \onlinecite{Abragam}). 
	The residual $\chi^{\rm V(3a/3b)}$ is plotted against $K^{\rm V(3a/3b)} = [K^{\rm V(3a)} + K^{\rm V(3b)}]/2$ in Fig. \ref{FigS2}(a). 
	The linearity yields the hyperfine coupling constant $A^{\rm s} = -9.5$ T/$\mu_{\rm B}$. 
	The crossing point with Eq. (S5) (the line with the positive slope) gives $K^{\rm orb} = 0.4\%$ and $\chi_{\rm orb}^{\rm V(3a/3b)} = 5 \times 10^{-5}$ emu/V(3a/3b)-mol. 
	It is natural to use the common $A^{\rm s}$ and $K^{\rm orb}$ in V(3a) and V(3b) with the similar orbital states, giving $\chi^{\rm V(3a)}$ and $\chi^{\rm V(3b)}$ from $K^{\rm V(3a)}$ and $K^{\rm V(3b)}$ [Fig. 3(a)]. 
	By subtracting $\chi^{\rm V(3a)}$ and $\chi^{\rm V(3b)}$ from $\chi$ below 140 K, we confirm that $\chi^{\rm V(1a/1b)}=[\chi^{\rm V(1a)} + \chi^{\rm V(1b)}]/2$ scales to $[K^{\rm V(1a)}+K^{\rm V(1b)}]/2$ with a reasonable $A^{\rm s} = -12$ T/$\mu_{\rm B}$, as shown in Fig. S2(b). 
	Using the common $A^{\rm s}$ for V(1a) and V(1b), $K^{\rm V(1a)}$ and $K^{\rm V(1b)}$ respectively give $\chi^{\rm V(1a)}$ and $\chi^{\rm V(1b)}$ [Fig. 3(a)]. 

	As for the metallic phase, three vanadium sites have negative Knight shifts, reflecting the predominant $K_{\rm iso}^{\rm s}$. 
	Using the common $A^{\rm s}$ and $\chi_{\rm orb}$ to the insulating phase, we obtain $\chi^{\rm V(3)}$ and $\chi^{\rm V(2)}$ from $K^{\rm V(3)}$ and $K^{\rm V(2)}$. 
	Finally, the subtraction of $\chi^{\rm V(3)}$ and $\chi^{\rm V(2)}$ from $\chi$ gives $\chi^{\rm V(1)}$. 

\section{B. Anisotropic Knight shifts} 
	The $d$ orbital occupation is reflected as the anisotropic hyperfine coupling. 
	The magnetic part of the nuclear spin Hamiltonian $\mathcal {H}_{\rm mag}$ is written as
	\begin{eqnarray}
	\mathcal {H}_{\rm mag} = -P[\kappa {\bf S\cdot I} + \xi {\bf S} \cdot {\bf {\sf q}} \cdot {\bf I}] - \gamma_N \hbar \bf H\cdot \alpha \cdot \bf I, 
	\end{eqnarray} 
	where ${\bf S}$ and ${\bf I}$ are electron and nuclear spin operators, respectively, $\hbar = h/2\pi$ is the Planck's constant, $\xi = \frac{2}{21}$, and $P$ represents the product of the Bohr magneton $\mu_{\rm B}$, the nuclear gyromagnetic ratio $\gamma_{\rm N}$, and the ionic radial average factor $\braket {r^{-3}} = 3.684$ a.u. for V$^{4+}$ ($3d^1$) \cite{Abragam}.
	The first term in Eq. (S6) represents the core polarization interaction giving the negative isotropic Knight shift proportional to the spin susceptibility $\chi_{\rm s}$ with a coefficient $\kappa$. 
	The second term arises from the dipole hyperfine interaction proportional to both $\chi_{\rm s}$ and the orbital polarization ${\sf q}$. 
	The components ${\sf q}_{ij}$ $(i, j = x, y, z)$ are expressed with the equivalent operators 
	\begin{eqnarray}
	\label{lij}
	{\sf q}_{ij} \equiv \frac{3}{2}(L_i L_j + L_j L_i ) - \delta_{ij}{\bf L}^2  
	\end{eqnarray} 
	with the total orbital angular momentum ${\bf L}$. 
	The third term corresponds to the temperature-independent Van-Vleck term arising from a cross term of ${\bf L}$ and the Zeeman interaction under magnetic field $H_\mu$ with the components $\alpha_{\mu \nu} = 4\mu_{\rm B}^2\braket {r^{-3}}\Lambda_{\mu \nu}$, where $\Lambda_{\mu \nu}$ is the second-order mixing probability of the ground and excited states. 

	In a $3d^1$ case, the principal components of ${\sf q}_{ij}$ are given by ${\sf q}_{zz} = -2{\sf q}_{xx} = -2{\sf q}_{yy} = 6$ for $d_{xy}$, ${\sf q}_{xx} = -2{\sf q}_{yy} = -2{\sf q}_{zz} = 6$ for $d_{yz}$, and ${\sf q}_{yy} = -2{\sf q}_{xx} = -2{\sf q}_{zz} = 6$ for $d_{zx}$ \cite{Abragam, Kiyama}. 
	For arbitrary occupations, ${\sf q}_{ij}$ can be written as a linear combination of three orbital components, $({\sf q}_{xx}, {\sf q}_{yy}, {\sf q}_{zz}) = 3(-\alpha+2\beta-\gamma, -\alpha-\beta+2\gamma, 2\alpha-\beta-\gamma)$, where $d_{xy}:d_{yz}: d_{zx} = \alpha: \beta: \gamma$ ($\alpha + \beta + \gamma = 1$). 
	In V$_6$O$_{13}$, the local coordinates $(x, y, z)$ are taken parallel to the crystal axes $(a, b, c^*)$ or $(c^\prime, b^\prime, a^{\prime *})$ defined in Fig. 1, , which are close to the V-O bond directions. 
	For metals, ${\sf q}_{ij}$ reflects the $d$ orbital anisotropy for the electrons at the Fermi level. 

\begin{table*}[h]
\begin{tabular}{ccccccccc}
\hline 
$T$ (K) &site&$K_X$(\%) &$K_Y$(\%) &$K_Z$(\%) &$\theta_X (^\circ)$&$\phi_X (^\circ)$&$\theta_Z (^\circ)$&$\phi_Z (^\circ)$
\\ \hline
170&V(1)& $-1.9$ &$-2.3$&$-0.9$& $-5$&11& $-84$&92\\ 
 &V(2)&$0.05$&$0.05$&$-0.60$&47&98&$-13$&1\\ 
 &V(3)&$-1.1$&$-1.2$&$-3.4$&$-2$&$-92$&$-72$&6\\ 
\hline
100&V(1a)&0.06&$-0.28$&0.06&75&10&87&80\\ 
&V(1b)&0.55&0.15&0.21&$-74$&23&$-10$&80\\
&V(3a)&$-1.0$&$-0.9$&$-2.6$&0&80&40&0\\
&V(3b)&$-1.5$&$-1.4$&$-3.8$&0&80&40&0\\
\hline
\end{tabular}
\caption{\label{TableS1}Principal components of the Knight shift, $(K_X, K_Y, K_Z)$, and the directions of the $X$ and $Z$ axes with respect to the crystal axes $(a, b, c^*)$ or $(c^\prime, b^\prime, a^{\prime *})$ at 170 and 100 K in V$_6$O$_{13}$, where $\theta$ is measured from $c*$ and $\phi$ from $a$ in the $ab$ plane. } 
\end{table*}

	By measuring the angular dependence of $K$ (Fig. 4) we obtained the principal components $(K_X, K_Y, K_Z)$, as displayed in Table S1. 
	We find that the principal axes are directed close to the crystal axes for V($i$) ($i$ = 1, 3, 1a, 1b, 3a, and 3b), in good agreement with the V-O bond directions $(x, y, z)$. 
	We can evaluate $K^{\rm orb}$ from the $K$-$\chi$ plots and extract $K^{\rm s}$ for each axis. 
	Then the Knight shift is decomposed into the isotropic part $K_{\rm iso}^{\rm s} = (K_X^{\rm s} + K_Y^{\rm s} + K_Z^{\rm s})/3$, the axial part $K_{\rm ax}^{\rm s} = (2K_Z^{\rm s} - K_X^{\rm s} - K_Y^{\rm s})/3$, and the in-plane asymmetric part $K_{\rm an}^{\rm s} = (K_X^{\rm s} - K_Y^{\rm s})/2$, which are expressed as $K_{\rm iso}^{\rm s} = -\kappa P \chi_s$, $K_{\rm ax}^{\rm s} = \frac{4}{7} P \chi_s (2\alpha - \beta -\gamma)$, and $K_{\rm an}^{\rm s} = \frac{6}{7} P\chi_s (-\beta + \gamma)$, respectively, for $x=X$, $y=Y$, and $z=Z$. 
	One can cancel out $\chi_s$ and extract the occupations by dividing $K_{\rm ax}^{\rm s}$ and $K_{\rm an}^{\rm s}$ by $K_{\rm iso}^{\rm s}$. 
	Using the typical $\kappa$ = 0.5 for vanadates \cite{Kiyama}, we obtain $K_{\rm ax}^{\rm s}/K_{\rm iso}^{\rm s} = \frac{4}{7}(2\alpha - \beta -\gamma)$ and $K_{\rm ax}^{\rm s}/K_{\rm iso}^{\rm s} = \frac{6}{7}(-\beta+\gamma)$. 
	The results for V(1), V(3), and V(3a/3b) are summarized in Table I. 

\section{C. Nuclear quadrupole frequency} 
	The nuclear quadrupole splitting $\delta \nu$ in presence of the electric field gradient (EFG) manifests the spatial $d$ electron distribution via the electric hyperfine interaction. 
	The electric hyperfine interaction $\mathcal {H}_{\rm el}$ between the $^{51}$V nucleus (the nuclear spin $I = 7/2$, the quadrupole moment $Q$ = $-0.05 \times 10^{-24}$ cm$^2$) and 3$d$ electrons is expressed as\cite{Kiyama} 
	\begin{eqnarray}
	\mathcal {H}_{\rm el} = \frac{2e^2Q}{I(2I-1)}\xi \braket {r^{-3}}[{\bf I}\cdot {\bf {\sf q}} \cdot {\bf I}]. 
	\end{eqnarray} 
	Here ${\sf q}$ has the same form as in the magnetic dipole hyperfine interaction in Eq. (S6) and reflects the anisotropic distribution of $3d$ electron clouds. 
	The nuclear quadrupole frequency $\nu_Q^d$ due to $3d$ electrons is given by 
	\begin{eqnarray}
	\nu_Q^d = \frac{e^2Q}{7hI(2I-1)}\braket {r^{-3}}{\sf q}_{\rm ZZ},  
	\end{eqnarray} 
	where ${\sf q}_{\rm ZZ}$ represents the principal components of the orbital polarization given by $3(-\alpha+2\beta-\gamma)$ for V(1a), $3(-\alpha-\beta+2\gamma)$ for V(1b), and $3(2\alpha-\beta-\gamma)$ for V(1), V(3), V(3a), and V(3b). 

	\begin{figure}
	\includegraphics[width=8cm]{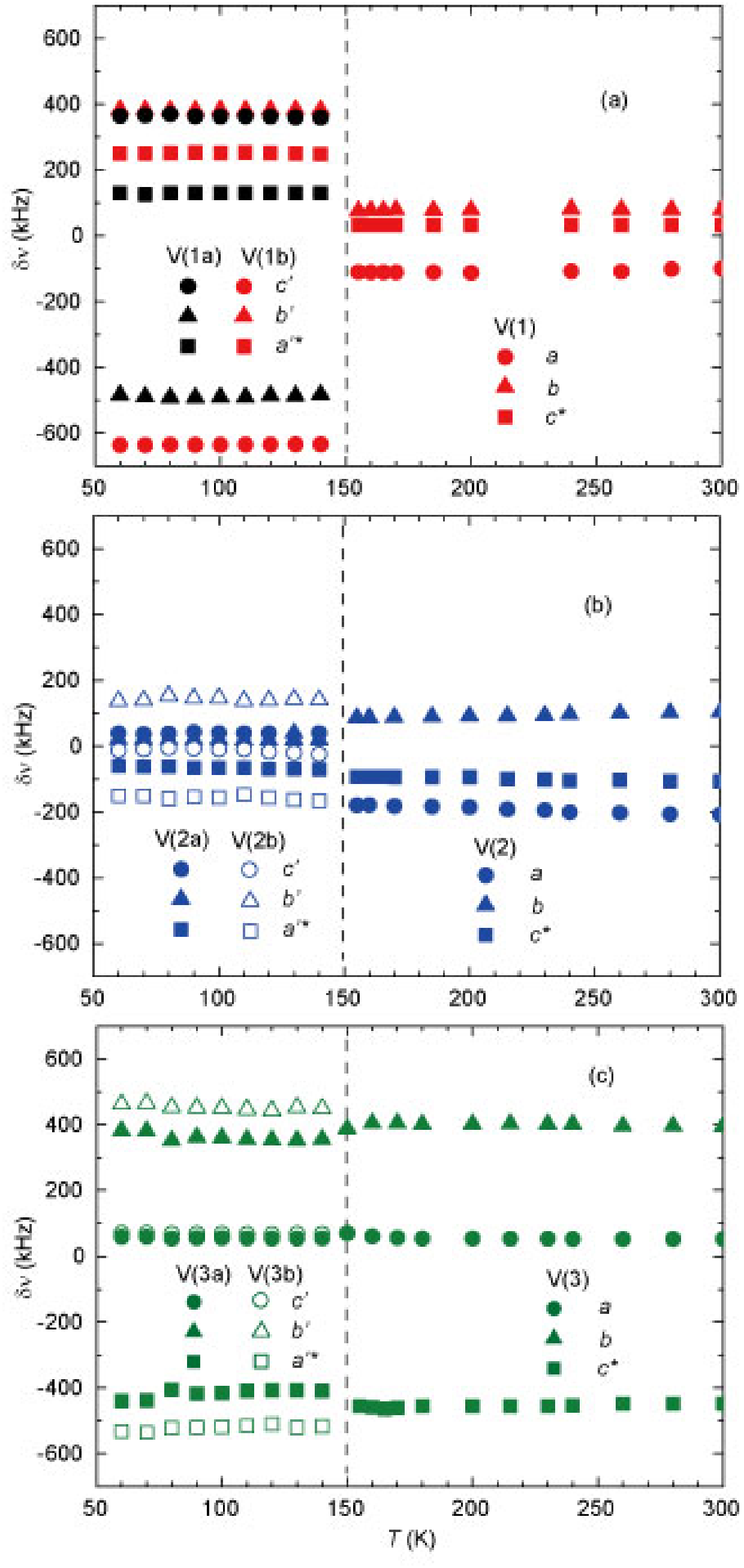}
	\caption{\label{FigS3} 
	Temperature dependence of $\delta \nu$ measured for $H_0$ parallel to the crystal axes in V$_6$O$_{13}$. 
	}
	\end{figure}
	
	The experimental results of $\delta \nu$ measured for the three crystal axes are shown in Fig. \ref{FigS3}. 
	The sign of $\delta \nu$ is selected so as to satisfy $\delta \nu_a + \delta \nu_b + \delta \nu_{c^*} = 0$ and agrees with the angular dependence of $K$.
	The maximum $|\delta \nu|$ gives the nuclear quadrupole frequency $\nu_Q$ proportional to the sum of the EFG from the outer ions and the $d$ electrons,  
	\begin{eqnarray}
	\nu_Q = (1 - \gamma_\infty)\nu_Q^{\rm out} + (1 - R)\nu_Q^d.  
	\end{eqnarray} 
	The outer ion contribution $\nu_Q^{\rm out}$ is enhanced by the core electrons with the Sternheimer antishielding factor, $\gamma_\infty$, whereas the $d$ electron contribution $\nu_Q^d$ is reduced with a shielding factor, $R \sim 0.35$ for vanadates \cite{Kiyama}. 

	The outer contribution to EFG, $V^{\rm out}_{mn}$ $(m,n = x, y, z)$, is expressed as 
	\begin{eqnarray}
	V^{\rm out}_{mn} = \sum_i q_i \Bigl( \frac{3m_i n_i- r_i^2 \delta_{mn}}{r_i^5}\Bigr), 
	\end{eqnarray} 
	with the charge on the $i$-th ion, $q_i$, and the distance between the V sites and the $i$-th ion, $r_i$. We evaluated $V^{\rm out}_{mn}$ with a point-charge approximation within a 200$\AA$ sphere using the crystal structure at 290 K for the metallic phase \cite{Dernier} and at 90 K for the insulating phase \cite{Howing}. 
	Here we postulated the valence state of V$^{4+}$ for V(1) [V(1a), V(1b)] and V(3) [V(3a), V(3b)], and V$^{5+}$ for V(2) [V(2a), V(2b)]. 
	Then the lattice contribution to the quadrupole splitting frequency, $\delta \nu^{\rm out}$, can be calculated as 
	\begin{eqnarray}
	\delta \nu_{mn}^{\rm out} = \frac{3eV^{\rm out}_{mn}}{h2I(2I - 1)}.  
	\end{eqnarray} 
	The results are shown in Table S2. 
	We can evaluate $\gamma_\infty$ as $\sim -5.5$ by comparing the averaged $\nu_Q \sim 240$ kHz (Table S3) and $\nu_Q^{\rm out} \sim 37$ kHz (Table S2) for V(2a) and V(2b) without on-site $d$ electrons. 
	Using the common $\gamma_\infty$ for other vanadium sites, we extract $\nu_Q^d$ for V(1) and V(3) with the on-site $d$ electron, as listed in Table S3. 
	From Eq. (S9), the orbital occupations are elucidated (Table S3). 
	The results are in qualitative agreement with those obtained from the Knight shift in Table I. 

\begin{table*}[h]
\begin{tabular}{ccccccccc}
\hline
$T$(K)&site&$\delta \nu_X$ (kHz)&$\delta \nu_Y$ (kHz)&$\delta \nu_Z$ (kHz)& $\theta_X (^\circ)$&$\phi_X (^\circ)$&$\theta_Z (^\circ)$&$\phi_Z (^\circ)$
\\ \hline
290&V(1)& 10.2 &12.1&$-22.3$& 90&0& 0&0\\ 
&V(2)&$-18.4$&$-19.3$&37.7&90&90& 4&1\\ 
&V(3)&$-23.5$&$-33.0$&56.6&90&90& 8&0\\ 
\hline
90&V(1a)&$-11.1$&$-24.8$&36.0&$78$&$-12$& $-12$&$-3$\\ 
&V(1b)&$-13.9$&$-20.1$&34.1&$92$&79& $79$&$-12$\\
&V(2a)&$-18.1$&$-20.7$&38.9&75&$-2$& $-15$& 0 \\ 
&V(2b)&$-13.8$&$-22.4$&36.3&$73$&$-4$& $-17$& 2 \\ 
&V(3a)&$-26.5$&$-31.1$&57.7&91&87& $-20$&2\\
&V(3b)&$-24.8$&$-31.2$&56.1&90&88& $-20$&1\\
\hline
\end{tabular}
\caption{\label{TableS3}Nuclear quadrupole splitting frequency calculated in a point charge approximation at 290 and 90 K in V$_6$O$_{13}$, and the directions of the $X$ and $Z$ axes. The $X$ direction is taken perpendicular to the $Z$ axis giving $\nu_{\rm Q}$.} 
\end{table*}

\begin{table*}[h]
\begin{tabular}{cccccccccccccc}
\hline 
$T$ (K)&site&$\delta \nu_X$ (kHz)&$\delta \nu_Y$ (kHz)&$\delta \nu_Z$ (kHz)& $\theta_X (^\circ)$&$\phi_X (^\circ)$&$\theta_Z (^\circ)$&$\phi_Z (^\circ)$&$\nu_Q^d$&$\eta ^d$&$\alpha$&$\beta$&$\gamma$
\\ \hline
170&V(1)& 33 &78&$-111$& $-5$&11& $-84$&92&183&1.0&0.24&0.33&0.43\\
&V(2)&163&84&$-247$&78&80&$-2$&43&-&-&-&-&-\\ 
&V(3)&404&56&$-460$&0&85&18&5&827&0.32&0.77&0.18&0.04\\ 
\hline
100&V(1a)&130&362&$-491$&75&10&87&80&725&0.18&0.11&0.18&0.71\\ 
&V(1b)&252&382&$-636$&$-74$&23&$-10$&$-80$&857&0.10&0.13&0.78&0.09\\
&V(2a)&96&$-187$&283&58&92&$-28$&29&-&-&-&-&-\\ 
&V(2b)&69&$-129$&198&58&92&$-28$&29&-&-&-&-&-\\ 
&V(3a)&60&340&$-440$&0&80&40&10&815&0.76&0.77&0.19&0.04\\
&V(3b)&68&480&$-540$&0&80&40&10&904&0.81&0.81&0.19&0.0\\
\hline
\end{tabular}
\caption{\label{TableS2}Principal components of the nuclear quadrupole splitting frequency, the directions of the principal axes, $\nu_Q^d$, $\eta ^d$, and the orbital components for the metallic (170 K) and insulating (100 K) phases in V$_6$O$_{13}$, where $\nu_{\rm Q}$ is defined as the maximum absolute value and the in-plane EFG anisotropy $\eta \equiv |\delta \nu_X - \delta \nu_Y|/\nu_{\rm Q}$. $X$ and $Y$ are taken as the directions perpendicular to the $Z$ axis giving $\nu_{\rm Q}$.} 
\end{table*}


\begin{thebibliography}{99}
\bibitem{note}
	  For the site assignment, we referred to directions of the displacements of V ions that determine the $d$ orbital state.
\end{thebibliography}
\end{document}